\documentclass[10pt,letterpaper,twocolumn]{article}
\usepackage[english]{babel}
\usepackage{graphicx}

\newcommand{\alt}{\mathbin{\lower 3pt\hbox
   {$\rlap{\raise 5pt\hbox{$\char'074$}}\mathchar"7218$}}}
\newcommand{\agt}{\mathbin{\lower 3pt\hbox
   {$\rlap{\raise 5pt\hbox{$\char'076$}}\mathchar"7218$}}}

\begin{document}

\setcounter{footnote}{0}
\setcounter{equation}{0}
\setcounter{figure}{0}
\setcounter{table}{0}

\title{\large\bf To separation of variables in the Fokker--Plank
equations }

\author{\small  I. M. Suslov \\
\small P.L.Kapitza Institute for Physical Problems,
119334 Moscow, Russia \\
\small E-mail: suslov@kapitza.ras.ru\\
{}\\
\parbox{150mm}{\footnotesize \,It is well-known, that for
separation of variables in the eigenvalue problem,
the corresponding operator should be represented as a sum
of operators depending on single variables. In the case of the
Fokker--Plank equations, separation of variables is possible
under essentially
%the more weak
weaker conditions.  } }

\date{}
\maketitle

%\textwidth 6.4 in
%\textheight 8.5 in

%\begin{document}
\setcounter{footnote}{0}
\setcounter{equation}{0}
\setcounter{figure}{0}
\setcounter{table}{0}
%\vspace*{2mm}

For separation of variables in the eigenvalue problem
$$
\hat L \, P(x,y)=\lambda \, P(x,y)
\eqno(1)
$$
the operator  $\hat L$ should be represented as a sum of
two operators $\hat L_x +\hat M_y$, depending only on $x$
and $y$ correspondingly.

Conditions for separation of variables in the Fokker--Plank
equations appear to be essentially weaker. For example, for the
equation describing the time evolution of the
probability distribution  $P\equiv P(x,y)$,
$$
\frac{\partial P}{\partial t}=
\left\{\vphantom{L^2} \hat L_{x,y} P \right\}'_x +
\left\{ \vphantom{L^2} \hat M_{y} P \right\}'_y \,,
\eqno(2)
$$
it is sufficient that the operator $\hat M_{y}$ in the last term
depends only on $y$, while the operator $\hat L_{x,y}$
remains arbitrary.  Indeed, setting $P=P(x) P(y)$ and dividing
by $P(x)$, one has
$$
-\frac{\partial P(y)}{\partial t} +
\left\{ \vphantom{L^2} \hat M_{y} P(y) \right\}'_y=
$$
$$=
\frac{P(y)}{P(x)} \frac{\partial P(x)}{\partial t}
-\frac{1}{P(x)}
\left\{\vphantom{L^2} \hat L_{x,y} P \right\}'_x
\,.
\eqno(3)
$$
The left-hand side is independent of $x$, and can be
considered as a certain function $F(y)$. Then
$$
P(y) \frac{\partial P(x)}{\partial t}
-\left\{\vphantom{L^2} \hat L_{x,y} P \right\}'_x
= F(y) P(x)
\eqno(4)
$$
and integration over $x$ gives $F(y)\equiv 0$,
since the left-hand side turns to zero, while the integral over
$P(x)$ is equal to unity due to normalization. As a result,
left-hand side and right-hand side of Eq.\,3 turn to zero
independently, and the equation for $P(y)$ is separated
$$
\frac{\partial P(y)}{\partial t} -
\left\{ \vphantom{L^2} \hat M_{y} P(y) \right\}'_y
=0 \,.
\eqno(5)
$$
On the other hand, integrating (3) over $y$, one has
$$
\frac{\partial P(x)}{\partial t} -
\left\{ \vphantom{L^2} \hat{\cal L}_{x} P(x) \right\}'_x
=0 \,,
\eqno(6)
$$
where
$$
 \hat{\cal L}_{x} = \int \hat L_{x,y} P(y) dy \,.
\eqno(7)
$$
The given considerations are very general and
applicable to any diffusion-type equation: the right-hand side
of the latter is always a sum of full derivatives, in order to
provide the conservation of probability. As a result,
the conditions for separation of variables appear to be
always weaker than for equation (1). In our opinion, this
fact should be mentioned in any courses of the mathematical
physics; unfortunately, it is not the case.

The separation of variables in the Fokker--Plank equations
was discussed in the comparatively new papers (e.g.
\cite{1,2,3}), but under rather restricted assumptions.
The equation of type (2) arises in the theory of 1D localization,
where it describes the evolution of the mutual distribution
$P(\rho,\psi)$ of the Landauer resistace $\rho$ and the
phase variable $\psi=\theta-\varphi$, where $\theta$
and $\varphi$  are phases entering the transfer matrix
(see Eq.28 in \cite{4} and the comments after it). Analogous
situation is expected in description of quasi-1D systems
in the framework of the generalized version
\cite{5} of the Dorokhov--Mello--Pereyra-Kumar equation
\cite{6,7}. It looks probable that analogous equations arose
in other fields of physics and in some cases the fact of
separation of variables was revealed by the corresponding
authors. However, the general character of this result was not
emphasized, and it remains unknown to wide audience.

\end{document}